\documentstyle[twocolumn,prl,aps,epsfig,amssymb,floats]{revtex}

\def\Journal#1#2#3#4{{#1} {\bf #2} (#4) #3}

\def\PLB{{\em Phys. Lett.}  B}
\def\PRL{\em Phys. Rev. Lett.}

\def\ra{\rightarrow}
\def \etv{E_T\!\!\!\!\!\!/~~} 
 
\def\@citexr[#1]#2{\if@filesw\immediate\write\@auxout{\string\citation{#2}}\fi
  \def\@citea{}\@cite{\@for\@citeb:=#2\do
    {\@citea\def\@citea{--\penalty\@m}\@ifundefined
       {b@\@citeb}{{\bf ?}\@warning
       {Citation `\@citeb' on page \thepage \space undefined}}%
\hbox{\csname b@\@citeb\endcsname}}}{#1}}
\catcode`@=12
\def\beq{\begin{equation}}
\def\eeq{\end{equation}}

\def\beqn{\begin{eqnarray}}
\def\eeqn{\end{eqnarray}}
\relax
\def\ba{\begin{array}}
\def\ea{\end{array}}

\def\beq{\begin{equation}}
\def\eeq{\end{equation}}
\def\bea{\begin{eqnarray}}
\def\eea{\end{eqnarray}}

\def\to{\rightarrow}

\def\[{\left[}
\def\]{\right]}

\def\U1em{{U(1)_{\rm em}}}
\def\ra{\rightarrow}

\def\sq2{\sqrt{2}}

\def\End{\end{document}}
\def \etv{E_T\!\!\!\!\!\!/~~} 

\def\thisday{October, 2000 and hep-ph/0010233~~} 
\begin{document}                                                              

\twocolumn[\hsize\textwidth\columnwidth\hsize\csname
@twocolumnfalse\endcsname

\title{Another Look at Charged Higgs Boson Production at LEP 
}  
\author{{\sc Shinya Kanemura$^{1}$, 
             Takashi Kasai$^{2,3}$, Guey-Lin Lin$^{4}$}, 
        {\sc Yasuhiro Okada$^{2,5}$, Jie-Jun Tseng$^{4}$}, 
        {\sc C.--P. Yuan}\,$^{1}$
}
\address{
$^1$~\emph{Physics and Astronomy Department, Michigan State University, 
           East Lansing, MI 48824-1116, USA} \\
$^2$~\emph{Theory Group, KEK, Tsukuba, 305-0801 Japan}\\
$^3$~\emph{Department of Accelerator Science,
The Graduate University of Advanced Studies,
Tsukuba,  305-0801 Japan}\\
$^4$~\emph{National Chiao Tung University, Hsinchu 300, Taiwan }\\
$^5$~\emph{Department of Particle and Nuclear Physics, 
The Graduate University of Advanced Studies,
Tsukuba,  305-0801 Japan} 
}

\date{\thisday}
\maketitle

\begin{abstract}
The current atmospheric and solar neutrino experimental data favors the 
bi-maximal mixing solution of the Zee-type neutrino mass matrix in which neutrino
masses are generated radiatively. This model requires the existence of a weak
singlet charged Higgs boson. While low energy data are unlikely to further
constrain the parameters of this model, the direct search of charged Higgs 
production at the CERN LEP experiments can provide useful
information on this mechanism of neutrino mass generation by analyzing their
data with electrons and/or muons (in contrast to taus or charms) in the 
final state with missing transverse energies. We also discuss the difference 
in the production rates of a weak singlet from a weak doublet charged Higgs 
boson pairs at LEP. 
\end{abstract}

\vskip1pc]

\setcounter{footnote}{0}
\renewcommand{\thefootnote}{\arabic{footnote}}


\section{Introduction}

Despite of the success of the Standard Model (SM), searching for new 
physics beyond the SM has always been one important part of the current
and future collider program. 
One such example is to detect the production of a charged Higgs boson at
the CERN LEP experiments. The most commonly studied new physics model 
other than the 
minimal supersymmetric Standard Model (MSSM) is the general two-Higgs
doublet model (THDM), which includes a pair of charged Higgs boson
($H^\pm$). Since $H^\pm$ is a component of the Higgs doublets, it
carries a nonvanishing weak quantum number. At electron-positron 
colliders, the production rate of the charged Higgs boson pair 
via the tree level process $ e^-e^+ \ra \gamma, Z \ra H^+ H^-$
is uniquely predicted. This is because the coupling of $\gamma$-$H^+$-$H^-$ 
is determined by the electric charge of $H^\pm$, and the coupling of 
$Z$-$H^+$-$H^-$ is determined by the weak charge (as well as electric charge) 
of the charged Higgs boson.
Furthermore, in the general THDM, the coupling of $H^\pm$ to fermions is 
determined by the Yukawa interaction. Usually, its strength is
proportional to the fermion masses, so that the $H^-$ 
predominantly decays into $\tau {\overline{\nu}_{\tau}}$ and $s {\bar c}$.
By examining the above decay modes, the LEP-II experimental groups have 
constrained the decay branching ratios of 
$H^- \ra \tau \bar{\nu_{\tau}}$ and/or $H^- \ra s {\bar c}$ 
as a function of the charged Higgs boson mass\cite{charged-search}.

In this letter, we propose to reexamine the LEP experimental data on
the detection of a charged Higgs boson to test new physics models 
in which a light weak singlet (not doublet) charged Higgs boson is present.
To clarify our discussions, we consider the Zee-model\cite{zee} as
an example.
From the observations of atmospheric and solar neutrinos, there are
increasing evidences for neutrino oscillations\cite{neuosc}.
If this is the correct interpretation, the SM has to be 
extended to incorporate the small neutrino masses suggested by
data. There have been several ideas proposed in the literature 
to generate small neutrino masses. 
The Zee-model is one of such attempts\cite{zee}, in which 
the three different flavor neutrinos are massless
at the tree level, and their small masses are generated
at the one-loop level.
For such a mass-generation mechanism to work, it is necessary to 
extend the Higgs sector of the SM to contain at least two weak-doublet 
scalar fields and one weak-singlet charged scalar field.
To generate a tiny neutrino mass through radiative corrections, this
singlet charged Higgs boson has to couple to leptons in different
families. 

According to the recent analysis in Ref.~\cite{Jarlskog_et_al,Jarlskog_may00}, 
the favored solution for having a bi-maximal mixing in the Zee-type
neutrino mass matrix is the MSW large angle solution, with the mass
relation $|m_{\nu_1}| \simeq |m_{\nu_2}| \gg |m_{\nu_3}|$, 
where $m_{\nu_i}$ ($i=1$-$3$) are the three neutrino masses.    
We shall show that, in such a case, 
the phenomenology of the singlet charged Higgs boson is 
completely different from that of the ordinary weak-doublet charged Higgs 
boson, especially in its decay pattern. 
In general, these two charged Higgs bosons will mix.
When the lighter charged Higgs boson ($S^\pm_2$) is the weak singlet, 
the dominant decay modes of $S^-_2$ are  
$e^- \etv$ and $\mu^- \etv$, while the $\tau^- \etv$ mode is highly
suppressed and the $s {\bar c}$ mode is not allowed at tree level.
The production rate of the $S^+_2 S^-_2$ pair at LEP-II is about 80\% of 
that of the $H^+ H^-$ pair production for the charged Higgs boson mass to 
be around 100 GeV. 
(Again, we denote $H^\pm$ as the charged Higgs boson in the usual THDM.) 
Therefore, the expected signal would escape the current analysis 
performed by the LEP-II experimental groups. 
To shed lights on this type of model 
(with a weak singlet charged Higgs boson), LEP-II collaborations 
should examine their data sample that contains $e^\pm$ and/or $\mu^\mp$ 
with the missing transverse energy ($\etv$).

\section{Zee Model}

The Zee-model requires a 
\( SU(2)_{L} \) singlet charged scalar field (\( \omega^{-} \)) in addition 
to two \( SU(2)_{L} \) doublet fields (\( \phi _{1},\phi _{2} \)).  
The Lagrangian can be written as: 
\begin{equation}
{\cal L}= {\cal L}_{kin}+{\cal L}_{Yukawa}+{\cal L}_{ll\omega }-V(\phi_{1},\phi_{2},\omega^{-}) \, ,
\end{equation}
where ${\cal L}_{kin}$ is the kinematic term and ${\cal L}_{Yukawa}$ is the usual 
Yukawa interaction term of the THDM.
The interactions among $\omega^-$ and leptons are defined by 
\begin{eqnarray}
{\cal L}_{ll\omega }=f_{ij}\overline{l_{i_{L}}}(i\tau_{2})(l_{j_{L}})^{C}
 \omega ^{-}+f_{ij}\overline{l_{i_{L}}}^{C}(i\tau_{2})l_{j_{L}}\omega ^{+} ,
\end{eqnarray}
 where $i$($=1,2,3)$ is the generation index, and
the left-handed lepton doublet $l_L$ is defined
 as 
\( \left( \begin{array}{c}\nu_\ell\\
                       \ell^- \end{array}\right)_L\).
The charge conjugation of a fermion field is defined as 
\( \psi ^{C}\equiv C\overline{\psi }^{T} \), where \( C \) is
the charge conjugation matrix (\( C^{-1}\gamma ^{\mu }C=-\gamma ^{\mu T } \)) 
with the super index $T$ indicating the transpose of a matrix.
The Higgs potential is given by 
\begin{eqnarray}
V(\phi _{1},\phi _{2},\omega^-) & = & m_{1}^{2}\left| \phi _{1}\right| ^{2}
+m_{2}^{2}\left| \phi _{2}\right| ^{2}
+m_{0}^{2}\left| \omega^{-}\right| ^{2} \nonumber \\ &&
\!\!\!\!\!\! \!\!\!\!\!\! \!\!\!\!\!\!  \!\!\!\!\!\!  
-m^{2}_{3}(\phi _{1}^{\dagger }\phi _{2}+
\phi ^{\dagger }_{2}\phi _{1})
-\mu \widetilde{\phi _{1}}^{T}i\tau _{2}\widetilde{\phi _{2}}\omega^{-}+
 h.c.\nonumber \\
 &  & \!\!\!\!\!\! \!\!\!\!\!\! \!\!\!\!\!\! \!\!\!\!\!\!  
+\frac{1}{2}\lambda _{1}\left| \phi _{1}\right| ^{4}+
 \frac{1}{2}\lambda _{2}\left| \phi _{2}\right| ^{4}+
 \lambda _{3}\left| \phi _{1}\right| ^{2}\left| \phi _{2}\right| ^{2}
 \nonumber \\
 &  & \!\!\!\!\!\! \!\!\!\!\!\! \!\!\!\!\!\! \!\!\!\!\!\!  
+\lambda _{4}\left| \phi ^{\dagger }_{1}\phi _{2}\right| ^{2}+
 \frac{\lambda _{5}}{2}\left[ \left( \phi ^{\dagger }_{1}\phi _{2}\right) ^{2}+
 \left( \phi ^{\dagger }_{2}\phi _{1}\right) ^{2}\right] \nonumber \\
 &  & \!\!\!\!\!\! \!\!\!\!\!\! \!\!\!\!\!\! \!\!\!\!\!\!  
 +\sigma _{1}\left| \omega^{-}\right| ^{2}\left| \phi _{1}\right| ^{2}+
 \sigma _{2}\left| \omega^{-}\right| ^{2}\left| \phi _{2}\right| ^{2}+
 \frac{1}{4}\sigma _{3}\left| \omega^{-}\right| ^{4}  , 
 \label{Higgs_potential} 
\end{eqnarray}
where
\( \phi _{m}=\left( \begin{array}{c}\phi _{m}^{0}\\
\phi _{m}^{-} \end{array}\right)\)
and 
\(\widetilde{\phi _{m}}\equiv \left( i\tau _{2}\right) \phi _{m}^{*} \)
with \( m=1,2  \). 
Without loss of generality, we can take the anti-symmetric matrix 
\( f_{ij} \) and the coupling \( \mu  \) to be real. 
In order to suppress the flavor changing neutral current (FCNC) 
at the tree level, a discrete symmetry 
(with $\phi_1 \ra +\phi_1$, $\phi_2 \ra -\phi_2$, $\omega^+ \ra + \omega^+$) 
is imposed, which is broken only softly 
by the \( m_{3}^{2} \) term and the $\mu$ term 
in Eq.~(\ref{Higgs_potential}). Under this discrete symmetry,    
there are two possible forms of the Yukawa interaction term, 
${\cal L}_{Yukawa}$, which are the same as that in the    
type-I and type-II THDM~\cite{hhg}.  
Hereafter, we shall refer to these two possible forms of ${\cal
L}_{Yukawa}$ as the type-I and type-II Zee-model, respectively.

We assume that the \( SU(2)_{L}\times U(1)_{Y} \) symmetry is broken 
to \( U(1)_{em} \) by the vacuum expectation values of \( \phi _{1}^0 \)
and \( \phi _{2}^0 \). 
The number of physical Higgs bosons are two CP-even Higgs bosons 
(\( H \),\( h \)), one CP-odd Higgs boson (\( A \)) and two pairs of 
charged Higgs bosons (\( S_{1} \),\( S_{2} \)). We adopt the convention, in
which  \( m_{H}>m_{h} \) and \( m_{S_{1}}>m_{S_{2}} \). 
In the basis where the two Higgs doublets are rotated by the vacuum angle 
\( \beta  \), with 
\( \tan\beta =\frac{\left\langle \phi _{2}^{0}\right\rangle }
{\left\langle \phi _{1}^{0}\right\rangle } \),
the mass matrix for the charged Higgs bosons is given by 
\begin{eqnarray}
\label{chargedmassmatrix}
M_{S}^{2}=\left[ \begin{array}{cc}
M^{2}-\frac{\lambda _{4}+\lambda _{5}}{2}v^{2} & 
\!\!\!\!\!\!\!\!\!\!\!\!\!\!\!\! -\frac{\mu v}{\sqrt{2}}\\
\!\!\!\!\!\!\!\!\!\!\!\!\!\!\!\! -\frac{\mu v}{\sqrt{2}} & 
\!\!\!\!\!\!\!\!\!\!\!\!\!\!\!\! m_{0}^{2}+\left( 
\frac{\sigma _{1}}{2}\cos ^{2}\beta +
\frac{\sigma _{2}}{2}\sin ^{2}\beta \right) v^{2}
\end{array}\right] \,,
\end{eqnarray}
where $M^2=m_3^2/(\sin \beta \cos \beta)$.
By diagonalizing this mass matrix, 
the original fields   
\( \phi _{1}^- \),  \( \phi _{2}^- \) and  \( \omega^{-} \) 
can be written as 
\begin{eqnarray}
\phi _{1}^{-} & = & G^{-}\cos \beta - 
(S_{1}^{-}\cos \chi -S_{2}^{-}\sin \chi )\sin \beta, \\
\phi _{2}^{-} & = & G^{-}\sin \beta +(S_{1}^{-}\cos \chi -
S_{2}^{-}\sin \chi )\cos \beta, \\
\omega^{-} & = & S_{1}^{-}\sin \chi +S_{2}^{-}\cos \chi \, ,
\end{eqnarray}
where \( G^{\pm } \) are the charged Nambu-Goldstone bosons, 
and the mixing angle \( \chi  \) characterizes the mixing between the
two mass eigenstates $S_{1}^{-}$ and $S_{2}^{-}$.

Although neutrinos are massless at the tree level, 
loop diagrams involving the charged Higgs bosons can generate 
the Majorana mass terms for all three neutrinos.
At the one-loop order, the neutrino mass matrix ($M_\nu$) is real 
and symmetric with vanishing diagonal elements 
in the basis where the charged lepton mass matrix  
is diagonal\cite{zee}.
The $(i,j)$ component of $M_\nu$ is given by  
\begin{equation}
\label{nmass2fij}
m_{ij}=f_{ij}(m_{e_{j}}^{2}-m_{e_{i}}^{2}) 
\frac{\mu \cot \beta }{16\pi ^{2}}\frac{1}{m_{S_{1}}^{2}-m_{S_{2}}^{2}}
\ln \frac{m_{S_{1}}^{2}}{m_{S_{2}}^{2}} \, ,
\end{equation}
where \( m_{e_{i}} \) \( (i=1,2,3) \) is the charged lepton mass for 
the type-I Zee-model. For the type-II model, 
$\cot \beta$ should be replaced by $\tan \beta$.

The phenomenological analysis of the above mass matrix 
in the Zee-model was performed in Refs.~\cite{Jarlskog_et_al,Jarlskog_may00}.
It was concluded that the bi-maximal mixing is the
only possibility to reconcile
the atmospheric and solar neutrino data\cite{Jarlskog_may00}. 
In terms of the three eigenvalues of 
the neutrino mass matrix, denoted as 
 \( m_{\nu _{1}} \), \( m_{\nu _{2}} \) and \( m_{\nu _{3}} \), 
the only possible pattern of the neutrino mass spectrum is 
\( \left| m_{\nu _{1}}\right| 
\simeq \left| m_{\nu _{2}}\right| \gg \left| m_{\nu _{3}}\right|  \), with 
\( m_{\nu _{1}}^{2}-m_{\nu _{3}}^{2}\simeq 
m_{\nu _{2}}^{2}-m_{\nu _{3}}^{2}=\Delta m_{atm}^{2} \),
and \( \left| \Delta m_{12}^{2}\right| =\left|
 m_{\nu _{1}}^{2}-m_{\nu _{2}}^{2}\right| =\Delta m_{solar}^{2} \),
where from the atmospheric neutrino data,
\( \Delta m_{atm}^{2}=O(10^{-3})\; {\rm eV}^{2} \), and 
 from the solar neutrino data,
\( \Delta m_{solar}^{2}=O(10^{-5})\; {\rm eV}^{2} \)
(the MSW large angle solution).
The above results imply 
\begin{eqnarray}
\left| \frac{f_{12}}{f_{13}}\right|  & \simeq  & 
\frac{m_{\tau }^{2}}{m_{\mu }^{2}}     \simeq    3 \times 10^2, 
\label{eq21}\\
\left| \frac{f_{13}}{f_{23}}\right|  & \simeq  & 
\frac{\sqrt{2}\Delta m_{atm}^{2}}{\Delta m_{solar}^{2}} 
\simeq     10^2. \label{eq22}
\end{eqnarray}
Therefore, the magnitudes of the three coupling constants should satisfy the
relation \( \left| f_{12}\right| \gg \left| f_{13}\right| \gg 
\left| f_{23}\right|  \),
which is crucial for studying the phenomenology of the singlet 
charged Higgs boson.

For a given value of the parameters $m_{S_1}$, $m_{S_2}$, $\tan \beta$, 
and $\mu$, the coupling $f_{ij}$ can be calculated from
Eq.~(\ref{nmass2fij}).
For example, for $m_{S_1}=500$\,GeV, $m_{S_2}=100$\,GeV, $\tan \beta=1$,
and $\mu=100$\,GeV, we obtain $|f_{12}| \sim 3 \times 10^{-4}$, assuming
$m_{12}= 3 \times 10^{-2}$\,eV. 
In the case that $m_{S_1}$ is large, the lighter charged Higgs boson  
$S^-_2$ is almost a weak singlet, i.e. the mixing angle $\chi$
approaches to zero.
For such values of $f_{ij}$ and $m_{S_2}$, it is unlikely to have an
observable effect from the Zee-model to the low energy data\cite{ng},
e.g., the muon life-time, the universality of tau decay into electron or 
muon, the rare decay of $\mu \to e \gamma$, the universality of 
$W$-boson decay into electron, muon or tau, 
and the decay width of $Z$ boson.
Since $|f_{ij}|$ are small, we do not expect 
a large rate in the 
lepton flavor violation decay of a light neutral Higgs boson, such as 
$h \ra \mu^\pm e^\mp$ (the largest one), $h \ra e^\pm \tau^\mp$, or 
$h \ra \mu^\pm \tau^\mp$ (the smallest one).
On the contrary, the decay width 
of $h \ra \gamma \gamma$ can change by about 20\%
as compared to the SM prediction, whose details will be
given elsewhere\cite{zeepaper}.
The direct search of a weak singlet charged Higgs boson
at the LEP experiments can further test this model as discussed in the 
following section.

\section{Decay and Production of $S^\pm_2$}

In the Zee-model, two kinds of charged Higgs bosons appear.   
If there is no mixing between them ($\chi=0$), the mass eigenstates 
$S_1^\pm$ and $S_2^\pm$ correspond to the THDM-like charged Higgs boson  
$H^\pm$ and the singlet Higgs boson $\omega^\pm$, respectively.  
Thus, the detection of $S_2^\pm$ can be a strong support of the Zee-model.
Here, we discuss how this extra charged boson $S_2^\pm$ can be 
detected at collider experiments.
For simplicity, we only consider cases with $\chi=0$.
Since  $m_{S_2}^{} < m_{S_1}^{}$, $\chi=0$ implies 
that the singlet charged Higgs is lighter than the doublet 
charged Higgs boson. 

The $S_2^-$ boson decays into 
a lepton pair $e_i^- \overline{\nu}_{e_j}^c$ with the coupling 
constant $f_{ij}$.   
The decay rate, 
$\Gamma^{S_2}_{ij}=\Gamma(S_2^- \to e_i^- \overline{\nu}_{e_j}^c)$,  
is calculated as 
\begin{eqnarray}
 \Gamma^{S_2}_{ij} = \frac{m_{S_2}^{}}{4\pi} \, f_{ij}\,^2 
             \left( 1 - \frac{m_{e_i}^2}{m_{S_2}^2} \right)^2,   
\end{eqnarray}
and the total decay width is given by 
\begin{eqnarray}
 \Gamma^{S_2}_{\rm total} = 
\sum_{i,j} \Gamma^{S_2}_{ij} .
\end{eqnarray}
Taking into account the hierarchy of $f_{ij}$ (c.f. Eqs. (\ref{eq21}) 
and (\ref{eq22}) ),  and assuming $|f_{12}| \sim 3 \times 10^{-4}$,  
we estimate the total decay width ($\Gamma^{S_2}_{\rm total}$) and 
the life time ($\tau$) as 
\begin{eqnarray}
\Gamma^{S_2}_{\rm total} 
&\sim& \Gamma^{S_2}_{12} + \Gamma^{S_2}_{21} 
 \sim  1.6 \; {\rm keV}, \;\;  \\
\tau &\sim& 1/\Gamma_{\rm total}^{S_2} \sim 10^{-18} {\rm sec}, 
\end{eqnarray}
for $m_{S_2}^{} \sim 100\;{\rm GeV}$.
Therefore, $S_2^\pm$ decays promptly after its production, and 
can be detected at collider experiments.

We note that $S_2^\pm$ only decays leptonically 
with branching ratios estimated to be 
\begin{eqnarray}
  B(S_2^- \to e^- \etv) &\sim& 0.5, \\
  B(S_2^- \to \mu^- \etv) &\sim& 0.5, \\
  B(S_2^- \to \tau^- \etv) &\sim& 
  {\cal O}\left( \frac{m_\mu^4}{m_\tau^4} \right) \sim 10^{-5},  
\end{eqnarray}
where we have used Eqs. (\ref{eq21}) and (\ref{eq22}).
Clearly, the branching ratio into the $\tau^- \etv$ mode 
is very small, so that it is not as useful for detecting $S_2^\pm$.  
This is different from the case of detecting the ordinary THDM-like 
charged Higgs boson, which preferentially decays into heavy fermion pairs 
(e.g. $\tau \nu$ and $cs$).

The main production channel for $S_2^\pm$ at the LEP-II experiment 
is the pair production process 
$e^+e^- \to S^+_2 S^-_2$, similar to the production of 
the THDM-like charged Higgs boson $S_1^+$.   
The matrix-element squares for the $S_i^+S_i^-$ production 
($i=1,2$) are calculated as  
\begin{eqnarray}
&& \left| {\cal M}
(e^-_{\footnotesize
\begin{array}{c}L\\R
\end{array}
}e^+_{\footnotesize
\begin{array}{c}R\\L
\end{array}
} \to S_i^+S_i^-) \right|^2 
=          \left\{  \frac{Q_e^{} e^2}{s}  \right.
\nonumber\\ && 
           \left.    - 
                  \frac{1}{c_W^2}                    
                 (I_{S_i}^3 - s_W^2 Q_{S_i}^{}) 
                  \frac{(I_e^3 - s_W^2 Q_e) g^2}{s - m_Z^2} \right\}^2 
s^2 \beta_{S_i}^2 \sin^2 \Theta,  
\end{eqnarray}
where $Q_e=-1$ and $I_e^3 = - \frac{1}{2}$ ($0$) for the 
incoming electron $e_L^-$ ($e_R^-$); 
$Q_{S_i}^{}=-1$ and $I_{S_i}^3 = - \frac{1}{2}$ ($0$) 
for $i=1$ ($2$);  
$\beta_{S_i^{}}=\sqrt{ 1 - 4 m_{S_i}^2/s}$, 
$s_W=\sin \theta_W$, $c_W = \cos \theta_W$, and 
$\Theta$ is the scattering angle of $S_i^-$ in the $e^+e^-$ 
center-of-mass (CM) frame whose energy is $\sqrt{s}$. 
For the other electron-positron helicity states  
($e^-_Le^+_L$ and $e^-_Re^+_R$), the cross sections are zero. 
Hence, the total cross section for the $S_2^+S_2^-$ pair production is 
\begin{eqnarray}
 \sigma(e^+e^- \to S^+_2 S^-_2) &=& 
     \frac{1}{96\pi} \,e^4  \beta^3_{S_2}  s   
     \left[ \left( \frac{1}{s} 
             + \frac{s_W^2}{c_W^2} \frac{1}{s - m_Z^2} \right)^2  
  \right. \nonumber\\
&& \!\!\!\!\!\!\!\!\!\!\!\!\!\!\!\!
\left.    +       \left\{ \frac{1}{s} 
              - \left( \frac{1}{2} - s_W^2 \right) 
  \frac{1}{c_W^2} \frac{1}{s - m_Z^2} \right\}^2  \right] . 
\end{eqnarray}
The ratio of the cross sections for $S^+_1 S^-_1$ and $S^+_2 S^-_2$ 
pair production,
$\sigma(e^+e^- \to S^+_2 S^-_2)/\sigma(e^+e^- \to S^+_1 S^-_1)$, 
is about $0.8$ at $\sqrt{s}=210$ GeV, 
when the masses of $S_1^\pm$ and $S_2^\pm$ are equal. 
Since both  cross sections have the same mass dependence, 
this ratio is a constant for each CM energy.    

The branching ratio of $S_2^- \to e_i^- \bar{\nu}^c_{e_j}$ 
with $e_i^- = e^-$ or $\mu^-$ is almost 100\%, so that we have 
$\sigma(e^+e^- \to S_2^+S_2^- \to \ell^+ \ell'^- \etv) 
\sim \sigma(e^+e^- \to S_2^+S_2^-)$, where 
$\ell^-$ or $\ell'^-$ represents $e^-$ and $\mu^-$ (but not $\tau^-$). 
Let us compare this with the cross section 
$\sigma(e^+e^- \to W^+W^- 
\to \ell^+ \ell'^- \etv) = 
 \sigma(e^+e^- \to W^+W^-) \cdot B(W^- \to \ell^- \etv)^2$, 
where 
$B(W^- \to \ell^- \etv)
 = B(W^- \to e^- \etv) + B(W^- \to \mu^- \etv) \sim 21\%$. 
As shown in Fig.~1,  the cross section 
$\sigma(e^+e^- \to S_2^+S_2^- \to \ell^+ \ell'^- \etv)$ 
is comparable with $\sigma(e^+e^- \to W^+W^- \to \ell^+ \ell'^- \etv)$. 
Therefore, we expect that, by carefully examining the 
$\ell^+ \ell'^- \etv$ events 
($\ell^+ \ell'^- = e^+e^-$, $e^\pm\mu^\mp$ and $\mu^+\mu^-$,  
in contrast to $\tau^+\tau^-$ for the $S_1^\pm$ case) in the LEP-II data, 
the lower bound on the mass of $S_2^\pm$ can be determined.
\begin{figure}[t]
{\par\centering \resizebox*{0.4\textwidth}
{!}{\includegraphics{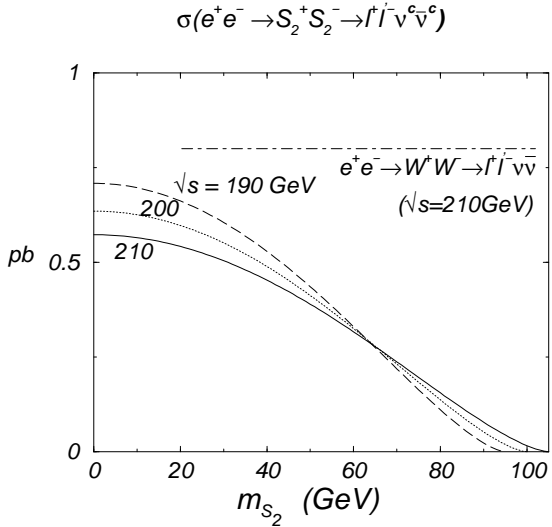}} \par}
\caption{
The cross section of the leptonic decay process 
$e^+e^- \to S_2^+S_2^- \to \ell^+\ell'^- \etv$ 
($\ell\, (\ell')= e$ and $\mu$) for $\sqrt{s}=190$, 200, 210 GeV. 
The process $e^+e^- \to W^+W^- 
\to \ell^+ \ell'^- \etv$ 
at $\sqrt{s}=210$ GeV is also shown, for comparison. }
\end{figure}

Finally, we comment on the $S_2^\pm$-production cross sections
at hadron colliders and future linear colliders (LC's),  
assuming no mixing between 
the charged Higgs bosons\cite{zeepaper}.
At hadron colliders, the dominant production mode 
is the pair production through the Drell-Yan-type process.    
The cross sections for $p \overline{p} \to S_2^+S_2^-$ 
at the Tevatron Run-II energy ($\sqrt{s}=2$ TeV) 
are about 24, 11, 2.2 fb for $m_{S_2}^{} = 80$, 100 and 150 GeV, respectively. 
At the LHC ($\sqrt{s} = 14$ TeV), the cross sections for 
$p p \to S_2^+S_2^-$ are 33, 3.3, 0.77 fb for 
$m_{S_2}^{} = 100$, 200 and 300 GeV, respectively.  
At future LC's, the $S_2^\pm$ bosons are produced mainly through 
$e^+e^- \rightarrow S_2^+S_2^-$ for $\sqrt{s}/2 > m_{S_2}$, and   
its cross section is about the same as that for 
the pair production of the ordinary weak doublet charged Higgs bosons in the 
THDM or the MSSM.

\section{Conclusion}

In summary, we pointed out that the phenomenology of the singlet charged 
Higgs boson can be completely different from that of the THDM-like 
charged Higgs boson.  For example, the singlet charged Higgs boson $S_2^-$ 
can decay into $e_i^- \overline{\nu}_{e_j}$ through the 
$f_{ij}$ couplings, where $e_i^-$ is $e^-$ or $\mu^-$. 
The decay branching ratio into the $\tau \nu$ mode is almost negligible 
for $|f_{12}| \gg |f_{13}| \gg |f_{23}|$, which results from 
fitting the neutrino oscillation data to the Zee-model mass matrix.  
On the other hand, the THDM-like charged Higgs boson $S_1^-$ 
decays mainly into 
$\tau^- \nu$ and $\overline{c} s$ 
through the usual Yukawa interactions.   
Therefore, to detect the singlet charged Higgs boson at LEP-II, 
one should examine the $\ell^+ \ell'^- \etv$ signal with 
$\ell^+ \ell'^- = e^+e^-$, $e^+\mu^-$, $\mu^+e^-$ or $\mu^+\mu^-$,  
in contrast to the usual detection modes of 
$\tau^+\tau^-\etv$, 
$\tau^\pm c s \etv$, etc.
The phenomenology of the Zee-model Higgs sector 
will be further discussed elsewhere\cite{zeepaper}.

\acknowledgments

We are grateful to the warm hospitality of the 
Center for Theoretical Sciences in Taiwan where part of this
work was completed. 
CPY would like to thank H.-J. He, J. Ng and W. Repko
for stimulating discussions.
SK was supported, in part, by the Alexander von Humboldt Foundation.
GLL and JJT were supported, in part, by the National Science Council of 
R.O.C. under the grant No NSC-89-2112-M-009-041; 
YO  was supported by the Grant-in-Aid of the
Ministry of Education, Science, Sports and Culture, Government 
of Japan (No.\ 09640381), 
Priority area ``Supersymmetry and Unified Theory of
Elementary Particles'' (No.\ 707), and ``Physics of CP Violation''
(No.\ 09246105); CPY is supported by
the National Science Foundation in the USA under the grant PHY-9802564.

\end{document}